\documentclass{article}
\usepackage[utf8]{inputenc}

\title{Quantum Computing: Towards Industry Reference Problems}
\author{Andre Luckow, Johannes Klepsch and Josef Pichlmeier\\
\\
BMW Group, Germany}
\date{}

\usepackage{graphicx}
\usepackage{hyperref}
\usepackage{url}
\usepackage{fullpage}
\usepackage{tablefootnote}
\usepackage{longtable}

\usepackage{xcolor}

\begin{document}

\maketitle

\begin{abstract}
The complexity is increasing rapidly in many areas of the automotive industry. The design of an automobile involves many different engineering disciplines, e.\,g., mechanical, electrical, and software engineering. The software of a vehicle comprises millions of lines of code. Further, the manufacturing, logistics, distribution, and sales of a vehicle are highly complex. There is an immense need for solving simulation problems, e.\,g., in battery chemistry, an essential enabler for technological advancements for electric vehicles. In all these domains, myriads of optimization, simulation, and machine learning problems arise. Quantum computing-based approaches promise to overcome some of the inherent scalability limitations of classical approaches. This article investigates quantum computing applications across the automotive value chain and identifies several high-value problems that will benefit from quantum-enhanced solutions.  
\end{abstract}

\section{Introduction}


The field of quantum computing is evolving rapidly. Progress in quantum hardware can be observed through the increasing number of qubits and growing quantum volume. The maximum quantum volume grew from 16 in 2019 to 64 in 2020~\cite{jurcevic2020demonstration}. The last decades were marked by the development of algorithms that offer a theoretical exponential speedup compared to classical methods, e.\,g., Shor's algorithm~\cite{365700}. Further, advantages of quantum solutions for different (non-practical) problems have been demonstrated, e.\,g., Google showed an advantage in a computational task involving random numbers~\cite{48651}. Chinese researchers created a quantum device for sampling bosons~\cite{Zhong1460}, demonstrating a problem that could not be solved in finite time on classical hardware. 


However, current quantum systems are limited to non-practical, small problems and are referred to as Noisy Intermediate Scale Quantum (NISQ) computers~\cite{Preskill2018}. Nevertheless, promising algorithms for these NISQ systems are emerging.
They utilize both classical and quantum hardware and are therefore often called hybrid algorithms. While most algorithms in this class show a similar working principle, the variety of problems that can be solved is increasing. Currently, this includes the simulation of chemical properties in small-scale molecules \cite{Kandala2017}, solving optimization problems \cite{farhi2014quantum}, performing machine learning tasks such as classification \cite{Havlek2019}, or finding solutions to differential equations \cite{kyriienko2020solving}.


The automotive industry's complexity makes the industry a prime candidate for quantum computing with many promising applications. Vehicles are highly complex products with many variants assembled in an international network of plants leading to many opportunities for optimization and machine learning methods. Further, the automotive industry's transition to electric vehicles makes battery chemistry simulation an important opportunity. This article aims to investigate applications in the automotive industry that can benefit from quantum computing.

Identifying high-value industry problems is crucial to guide progress in the field of quantum computing~\cite{acatech}. In this paper, we propose the usage of industry reference problems to develop benchmarks. For this purpose, we investigate industry applications that (i) are limited by the current state of optimization, simulation, and machine learning, (ii) have potential quantum-enhanced solutions. We postulate that solutions to these reference problems help guide quantum computing toward commercially valuable solutions.

This paper is structured as follows: Section~\ref{sec:apps} investigates applications in both the product and industry 4.0 domain. We continue with a discussion of the methodology for the development of industry benchmarks in section~\ref{sec:approach}. Based on the described applications, we identify potential reference problems and provide an in-depth description for two problems: robot path planning and vehicle configuration optimization.

\section{Applications in the Automotive Value Chain}
\label{sec:apps}


We discuss potential application areas for quantum computing as an enabler for new product capabilities and Industry 4.0. Table~\ref{tab:automotive_apps} provides an overview of common application scenarios. In the following, we discuss selected applications in detail.

\begin{table}[t]
    \centering
    \scriptsize
    \begin{tabular}{|p{0.25cm}|p{3.3cm}|p{5.2cm}|p{6cm}|}\hline
         &\textbf{Application Scenario} &\textbf{Variants} &\textbf{Description}  \\\hline
         1& Robotic Path Optimization & Bodyshop, paintshop, assembly, logistics & Hundreds of robots per plant in particular in body and paint shops~\cite{9108020,streif2020beating,DBLP:conf/kivs/MehtaMW17} \\\hline

         2& Vehicle Configurations & Vehicle options optimizations, crash relevant component layout, cell layout for heat optimization, parts demand calculation, seat position path optimization & Modern vehicles contain many independent but interconnected subsystems for which an optimal combination needs to be derived for reasons including emission and endurance testing \cite{Astesana2010} \\\hline
         
         3 & System Verification & Verification of automated systems, software testing  & Testing of connected subsystems that are designed and described through diverse models to ensure the safety of cyber-physical systems \cite{DaszczukMNW17, Kyoung-Dae2012} \\ \hline
         
         4 & Route Optimization &Logistics, fleets, car sharing, routing & In automated driving and  on-demand-mobility, finding optimal paths is crucial \cite{doi:10.1080/00423114.2014.939094,roch_capacitated_2019} \\\hline
         
        5 & Placement \& Distribution Problems  & Charging station placement, on-demand vehicle distribution &  Complex, non-linear problems aiming to optimize the geographic distribution of assets \cite{deb_nature-inspired_2021}  \\\hline

         6 & Strategic Planning & Volume planning, plant strategy, plant/model allocation & Highly complex, long term corporate planning (see Figure~\ref{fig:optimization_in_production}) \\\hline
         
         7 & Tactical Planning & Design of work stations in assembly, workforce planning, rework minimization & Various types of production failures, e.\,g., technical and human errors, lead to rework, and thus, higher costs~\cite{LIM2017469} \\\hline
          
         8 & Operational Planning & Workforce allocation, line balancing, shift scheduling, vehicle sequencing  & Highly customizable products lead to complex production lines with varying cycle times \cite{Bihlmaier2009}) \\\hline
         
         9 & Portfolio optimization & Feature selection in credit scoring, arbitrage opportunities, trading trajectories, risk analysis, Pricing of financial derivatives & Selection of optimal asset distribution considering various objectives, e.\,g., expected return, volatility \cite{ORUS2019100028}\\\hline

         10 & Nanoscale Functional Materials Development & Battery simulation, hydrogen simulation, corrosion inhibitors, material science for car body design  &Understanding molecular dynamics and electronic structure, simulating surface reactions in battery materials \cite{Elfving2020SimulatingQC} \\\hline

         
         

         11 & Engineering \& Design & Aeroacoustic simulation, component layout for cooling systems, airborne noise optimization, CFD   & Solving sets of differential equations is a key element in the development process of any vehicle \cite{yildiz}  \\\hline         
         
         12 & Computer Vision   & Visual inspection in manufacturing, object detection in automated driving & Improving the classification of $Nd$ data through representation in high dimensional Hilbert space \\\hline
         
    \end{tabular}
    \caption{\textbf{Automotive Applications:} A wide variety of optimization, simulation, and machine learning problems exist in the automotive value chain, indicating the potentially large impact of the technology on the industry.} 
    
    \label{tab:automotive_apps}
\end{table}

\subsection{Product and Customer}


Computing and software play a vital role in providing innovative capabilities, e.\,g., electric mobility, connected services, and automated driving~\cite{charette2009car}. Advances in quantum computing may enable significantly improved products and services in these domains. 
A critical area is research in battery materials that offer improved properties, considering, e.\,g., energy density, weight, and safety. Lithium metal anodes could, e.\,g., increase the energy density drastically~\cite{SCHNELL2018160}, require, however, other types of electrolytes, which are not available on industry-scale (\#10 in Table~\ref{tab:automotive_apps}). 
Simulations are critical for understanding potential materials and chemical reactions. However, classical algorithms for accurately simulating the behavior of molecules are computationally very costly. Quantum computing promises to push the boundaries of what is technically feasible in chemistry simulation~\cite{Elfving2020SimulatingQC}.

Another challenge for electric mobility is establishing and managing charging infrastructures (\#5 in Table~\ref{tab:automotive_apps}). The energy grid must be optimized to support the needs of the growing number of electrified vehicles, e.\,g., the increased and highly dynamic charging demands. Quantum-enhanced algorithms promise to optimize charging station positions, grid utilization and charging times (see~\cite{deb_nature-inspired_2021} for a review). 

Further, optimization problems arise in many connected services, e.\,g., routing, traffic optimization (\#4 in Table~\ref{tab:automotive_apps}). Problems in which the shortest possible route across a set of points has to be found can be solved using quantum algorithms to improve traffic predictions, routing, ride-sharing, and other types of on-demand-mobility~\cite{PhysRevA.95.032323}. 

\subsection{Industry 4.0}

Industry 4.0~\cite{dastbaz2019industry} envisions digital engineering and factories enabled by technologies, such as the IoT, cloud computing, augmented reality, 3-D printing, robotics, and artificial intelligence. Using these technologies enables shorter development time, highly customizable products (batch size ``one''), shorter production times, higher quality, and lower costs. The usage of optimization and machine learning methods is essential to achieve these goals (\#1, 7 \& 8 in Table~\ref{tab:automotive_apps}). 

Further, there are ample opportunities for using quantum computing to improve processes, e.\,g., in manufacturing, logistics, and design (\#1, 4 \& 11 in Table~\ref{tab:automotive_apps}). Designing a complex product such as a vehicle is challenging as designs have to meet various criteria, such as cost, safety, regulation, space, aerodynamic, weight, customizability, durability, manufacturing methods, and aesthetics. The highly dynamic market environment demands new approaches for solving engineering challenges. For example, generative design methods~\cite{bmw_gen_design} can improve the design and creative process by minimizing trial-and-error techniques.  Generative design methods are heavily reliant on AI and optimization approaches (\#11 in Table~\ref{tab:automotive_apps}).

Manufacturing automobiles is a complex challenge: not only is every vehicle a complex product on its own, but the high number of variants is also driving complexity to unprecedented scales~\cite{10.1145/2791060.2791071}. Combined with the large-scale volume of a typical manufacturer of ten thousands vehicles/day, this provides ample opportunities for applying quantum-based machine learning, simulation, and optimization methods.


\begin{figure}
    \includegraphics[width=\textwidth]{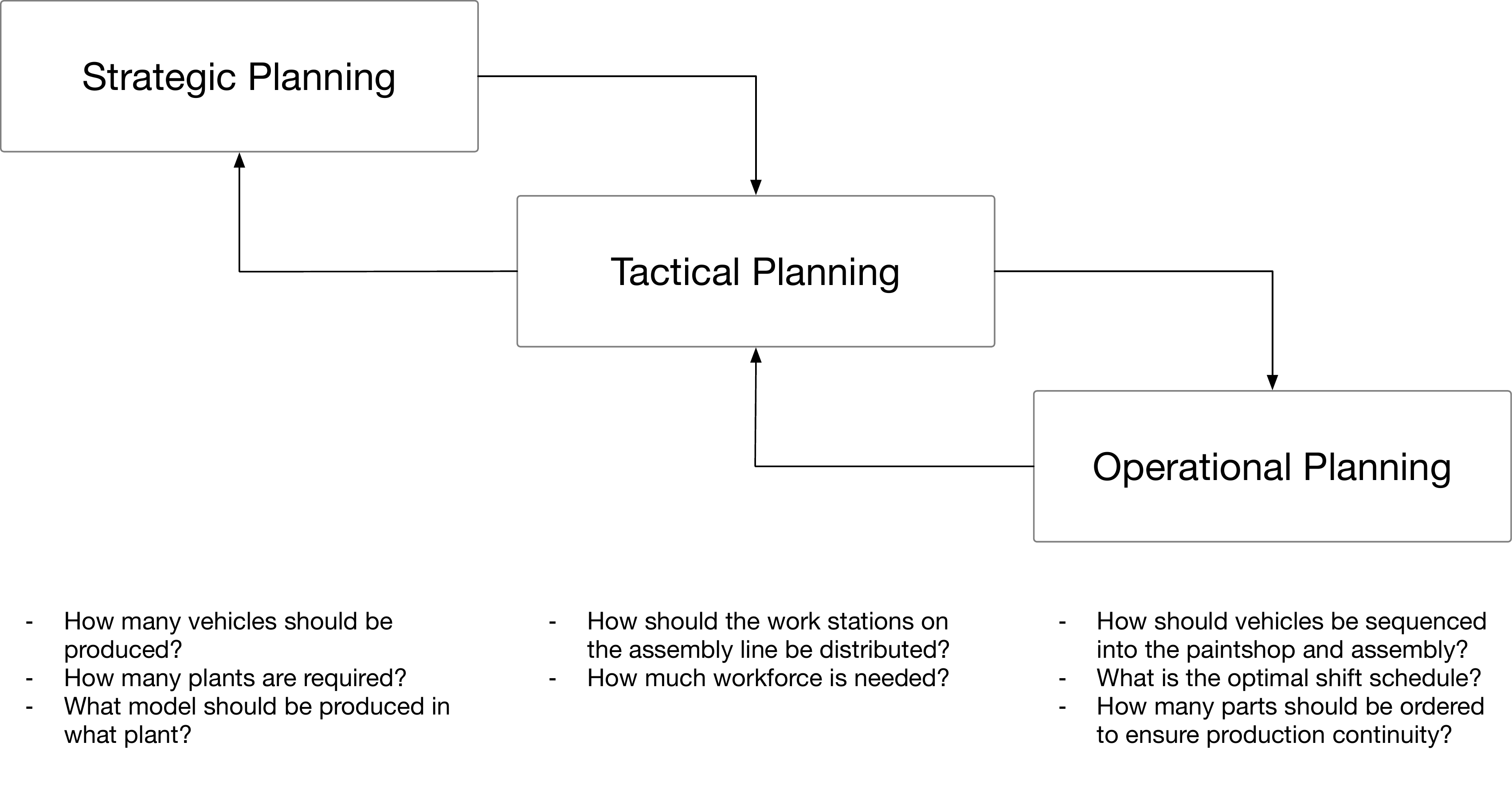}
    \caption{\textbf{Multi-Level Planning in the Automotive Industry:} Planning requires the consideration of different time-scales and granularities. Alignment within and across these layers is crucial to improve quality but results in added complexity.~\label{fig:optimization_in_production}}
\end{figure}

An important problem in the automotive value chain is planning across different time scales and granularities. Figure~\ref{fig:optimization_in_production} illustrates three planning dimensions: strategic, tactical, and operational planning (\#7, 8 \& 9 in Table~\ref{tab:automotive_apps}). Planning is highly challenging, involving many stakeholders and data across the company and the supply chain. Both horizontal alignment of data, models, and decisions across stakeholders on each level and vertical alignment is challenging for technical and business reasons.

The availability of data across the automotive value chain is increasing, as evidenced by the Automotive Alliance, which seeks to establish a collaborative industry platform for exchanging data across the automotive value chain~\cite{ganser}. This data provides the foundation for value chain optimizations and the ability to support long-term strategic decisions, such as allocating vehicle models to plants or the product mix considering complex constraints, such as capacities, logistics, labor costs, and customs.

Tactical decisions include issues related to the plant design (e.\,g., line balancing~\cite{Pearce2015}) and workforce allocation (e.\,g., shift scheduling). Many aspects of production are suitable for optimizing the operational level,  e.\,g., robot path optimization~\cite{9108020} and paint sequencing~\cite{streif2020beating}.


\section{Towards Industry Reference Problems}
\label{sec:approach}
\label{sec:benchmarking}



All layers of the quantum computing stack are evolving rapidly. At the same time, a gap between high-impact automotive applications and quantum solutions exists. This section discusses our methodology, identifies potential quantum solutions for industrial applications, and describes two potential reference problems.



\subsection{Methodology}

\begin{figure}[t]
    \centering
    \includegraphics[width=0.8\textwidth]{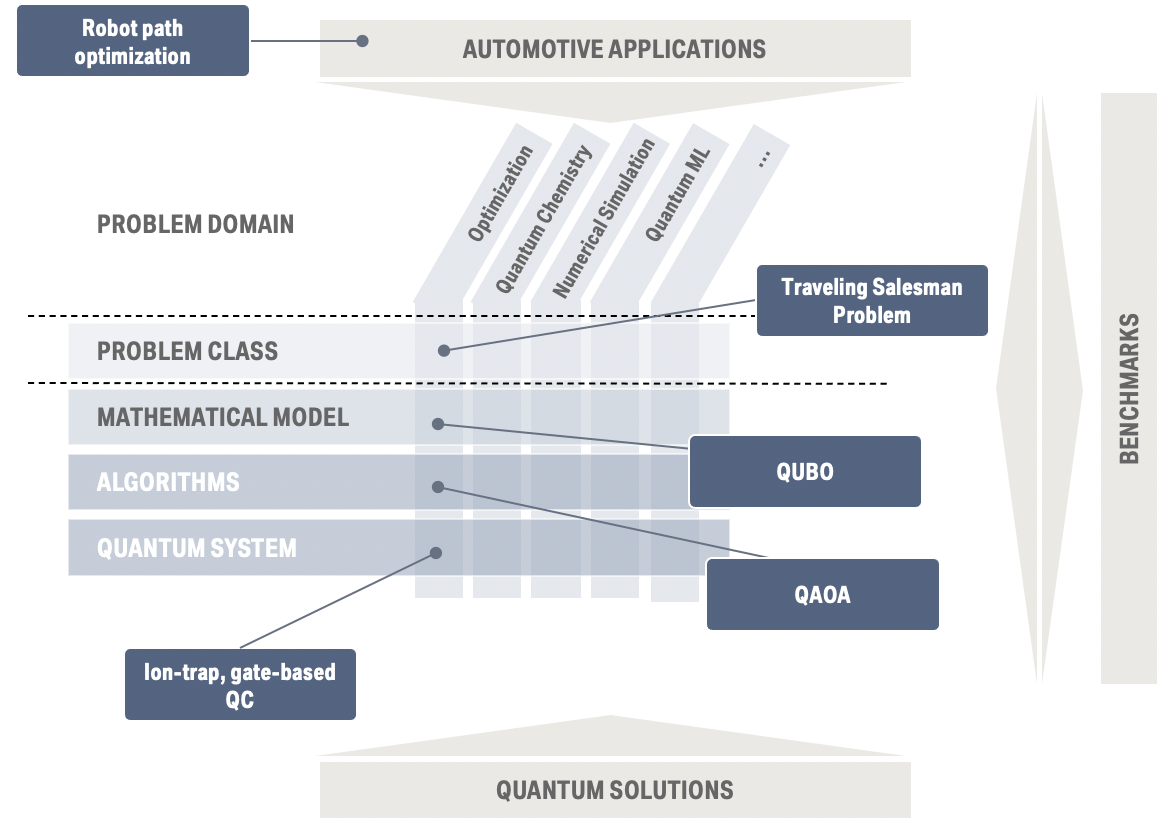}
    \caption{\textbf{Methodology:} We use a three-pronged approach: (i) identifying high-impact industry applications, (ii) deriving potential applications based on current quantum solutions, and (iii) using and developing benchmarks. \label{fig:approach2}
    }
\end{figure}

    Figure~\ref{fig:approach2} illustrates our methodology for identifying high-value, commercially relevant industry problems. We investigate (i) available quantum solutions, e.\,g., low-level quantum hardware and software, and (ii)  high-level application problems derived from real-world industry applications. Both approaches complement each other: progress on low-level challenges, e.\,g., hardware and algorithms, can inspire new industrial applications. High-impact industrial problems can accelerate progress on quantum solutions due to the high commercial relevance. Standardized benchmarks form the third pillar for comparing system designs and guiding progress.

We identified the following layers: problem domain, problem class, mathematical formulation, algorithms and quantum systems. We focus on four problem domains: optimization, quantum chemistry, numerical simulation and machine learning. For each domain, we identified different problem classes, e.\,g., traveling salesman, knapsack, graph partitioning for optimization. The mathematical formulation provides  a formal description of the underlying problem that enables assessment of the complexity, the required input, and the expected output. Often, different types of algorithms are suitable for solving a problem based on a given mathematical formulation. Commonly, we differentiate between quantum, hybrid, and quantum-inspired algorithms. The quantum system encompasses the software and hardware required to execute an algorithm, i.\,e., qubits, control, connectivity, and software. 


Quantum benchmarks are still in their infancy. While various performance characterizations of quantum systems have been conducted, they often lack comparability. While this is an important first step to understand current systems, it is essential to standardize evaluation processes and develop accepted community benchmarks. Benchmarking is an instrumental method to achieve this goal. Designing a good benchmark is a challenging task and requires careful consideration of often conflicting objectives, e.\,g., trading off different types of hardware, algorithms, and application characteristics. At the same time, the breadth of potential applications is high. A further complicating factor is a high uncertainty regarding the potential impact of the different factors on a potential quantum advantage.

\subsection{Background and Related Work}

A benchmark is a standardized workload, i.\,e., a program and a set of inputs, that is used to compare computer systems~\cite{ferrari1978computer}. Benchmarks are essential to compare systems, applications, algorithm innovation and steer progress. An example is ImageNet~\cite{5206848}, which motivated the creation of AlexNet and led to deep learning. We postulate that industry reference problems provide a good foundation for application benchmarks.

Benchmarks exist in different granularities. A kernel represents the central part of a program that contains the most time-consuming calculations~\cite{6448963}. Micro-benchmarks investigate a well-defined, narrow aspect of a larger system~\cite{Poggi2019}.  A benchmark suite comprises multiple benchmarks typically to study different aspects (e.\,g., application domains).

Benchmarking quantum systems is challenging due to various reasons, e.\,g., the complexity induced by various error types, e.\,g., state preparation, coherence and measurement errors, and the need to handle rapidly evolving hardware, software, and algorithms. Randomized benchmarks aim to address this challenge~\cite{Magesan_2011,MICHIELSEN201744}. The Quantum Volume benchmark emerged as a standard benchmark that captures manifold aspects of quantum computers, such as the number of qubits, gate fidelity, and error rates, into a single metric~\cite{PhysRevA.100.032328}. 

While low-level benchmarks aid the comparison of quantum systems, they have limited applicability to real-world applications. In particular considering that NISQ-era algorithms, such as hybrid algorithms, are highly dependent on specific hardware characteristics, e.\,g., specific noise characteristics, qubit connectivity, and other gate properties. Both the problem structure of the application and hardware characteristics are not sufficiently captured by the Quantum Volume metric.

Some application benchmarks have been proposed: Perdomo et al.~\cite{Perdomo_Ortiz_2019} utilize a randomized application benchmark for combinational circuit fault diagnostics, a SAT problem, to assess classical, quantum-inspired, and annealing optimization algorithms. Willsch et al.~\cite{willsch_benchmarking_2020} investigate the QAOA algorithm's performance on the weighted Max-Cut and 2-satisfiability problems. Q-score~\cite{atosqscore2020} is a proposed benchmark that measures a quantum processor's performance when solving standard combinatorial optimization problems. The score is calculated based on the maximum number of variables that a quantum solution can optimize. 



While these benchmarks provide immense value in measuring progress, they focus on singular aspects of quantum systems, e.\,g., the quantum volume is a singular metric incorporating many aspects. Quantum volume particularly emphasizes gate fidelity while weighing the number of qubits less~\cite{aaronson2020}. Thus, it is essential to develop benchmark suites that provide a broad perspective on the state of quantum computing. In particular,  application benchmarks based on industry problems can guide and drive progress towards commercially relevant problems.

\subsection{Understanding Automotive Applications}

Table~\ref{tab:benchmark_kernels} investigates problem domains, classes, mathematical formulation as well as quantum-based approaches for addressing these. Further, we map these to the described industry applications. The majority of applications are in the optimization domain, but notable applications in quantum chemistry, numerical simulations, and machine learning exist.

\begin{table}[h]
    \centering
    \scriptsize
    \begin{tabular}{|p{2.8cm}|p{3cm}|p{3cm}|p{3.5cm}|p{2cm}|}\hline
         \textbf{Problem Domain}  &\textbf{Problem Class} &\textbf{Mathematical Formulation} &\textbf{Algorithm} &\textbf{Application (ref. Table~\ref{tab:automotive_apps})} \\\hline
         Optimization & TSP, SAT, Max-Cut, partitioning, maximum independent set problem & QUBO, PUBO, Graph & QAOA \cite{farhi2014quantum} , Quantum Adiabatic Algorithm \cite{Tameem2016}, Grover Adaptive Search \cite{Austin2019} & [1,2,3,4,5,6,7,8,9] \\\hline
         
         Quantum Chemistry & Molecular dynamics  & Jorda-Wigner, Bravyi-Kitaev Superfast, Parity encoding  & QPE \cite{Abrams1997}, VQE \cite{Kandala2017} &[10]\\\hline
         
         Numerical Simulation &Computer fluid dynamics (CFD), crash simulation, automotive structure design  & Sets of differential equations, system of linear equations  & Harrow, Hassidim and Lloyd (HHL)~\cite{PhysRevLett.103.150502}, differential quantum circuits (DQC)~\cite{kyriienko2020solving} & [11] \\\hline
         
         Machine Learning  & Vision, natural language processing & Reproducing Kernel Hilbert spaces &  Variational Quantum Classifier \cite{Havlek2019}, Kernel Methods~\cite{Schuld2021}, Quantum Neural Networks & [12] \\\hline
    \end{tabular}
    \caption{\textbf{Application of Quantum Approaches to Automotive Applications:} Various mid- and long-term approaches for addressing specific optimization, simulation, and machine learning problems exist.
    }
    \label{tab:benchmark_kernels}
\end{table}

Quantum chemistry relies heavily on the simulation of Hamiltonian dynamics. By describing a system in the form of a many-body wave function, quantum computers can efficiently simulate their time evolution~\cite{Elfving2020HowWill}. A molecular Hamiltonian can be mapped onto a gate-based quantum computer using the mathematical transformations depicted in Table~\ref{tab:benchmark_kernels}. Different algorithms to simulate the evolution of the many-body Hamiltonian exist. The Variational Quantum Eigensolver (VQE) is designed for NISQ devices and tries to find a ground state of the Hamiltonian by using a classical optimizer in combination with a quantum computer. Quantum Phase Estimation (QPE) is an algorithm frequently used in many areas of quantum computing. It allows to derive the eigenstates and therefore eigenenergies of any Hamiltonian. However, it uses a circuit with high depth and therefore requires error-corrected quantum computers~\cite{Elfving2020HowWill}.

Many science and engineering problems, e.\,g., crash simulations and fluid dynamics, are expressed through partial differential equations. One approach is to use quantum systems as an accelerator in a finite element method~\cite{Montanaro2015}, e.\,g., by solving the corresponding system of linear equations using a variant of the HHL (Harrow, Hassidim, and Lloyd) algorithm~\cite{PhysRevLett.103.150502}. While the HHL algorithm promises a quantum speedup, it is not feasible on NISQ devices. However, promising hybrid and quantum-inspired approaches have been proposed, e.\,g., differential quantum circuits~\cite{Kyriienko2020} and tensor networks.

Many quantum machine learning approaches utilize a higher dimensional space by encoding classical data as quantum states \cite{Schuld2021}. Quantum circuits for encoding classical data are an active area of research. The goal is to achieve a mapping that is hard to simulate by classical computers~\cite{Havlek2019}. The resulting state is then processed by a parameterized quantum circuit (PQC), allowing exploration of the created space~\cite{Hubregtsen2020}. In the end, the states are measured, and a label is associated with the outcome. By training the parameters of the PQC, the probability that the correct label is assigned to unclassified data can be increased. This approach is sometimes also referred to as Quantum Neural Networks (QNN)~\cite{Killoran2018}.


The solution space for addressing the identified industry challenges is vast: for each problem, various mathematical formulations and algorithms suitable for different types of quantum hardware and system exist. As hardware and software co-design driven by these problems will be critical in the NISQ era, it is essential to establish standardized reference problems to guide designs and evaluate the trade-off between these solutions.

\subsection{From Industry Problem to Reference Benchmark}

Quantum solutions are advancing on all levels, on hardware, software, algorithm and mathematical formulation level. To support a structured investigation and future benchmarking, we propose the usage of reference problems. A reference problem comprises a description including an assessment of the business value, an analysis of the problem class, potential mathematical formulations, and quantum solutions. 

Unlike classical systems, current quantum systems are restricted to small (non-practical) problem sizes. Thus, it is essential to provide simplifications of the problems: They should be simple enough to explore current quantum systems, but still capture the essence of the application. However, they also need to support varying degrees of difficulties to accommodate future advances. Another challenge is assessing the quality of the application's output, particularly considering the stochastic nature of quantum computing. For example, while it is possible to assess the computational complexity for optimization problems based on the number of variables, it is difficult to verify the output, particularly as problems become intractable for classical machines.

The abstraction of applications onto problem classes and reference problems enables horizontal, vertical, and cross-industry collaboration. By abstracting high-value applications to problem classes of interest to a broad research and industry community, advances can be accelerated. Simultaneously, the defined problem classes can guide the hardware/software co-design (vertical collaboration), aligning quantum application, software, and hardware developers.  In the future, these problems serve as a basis for the investigation and the experimental evaluation of quantum-enhanced approaches, e.\,g., assessing the impact of noisy qubits, gate fidelities, and coherence times. We continue with a case study of two reference problems from the domain of optimization.


\subsubsection{Robot Path Optimization}
\label{sec:case_study}

\emph{Problem description:} In modern vehicle manufacturing, robots take on a significant workload, including performing welding jobs, sealing welding joints, or applying paint to the car body \cite{9108020}. While the robot's tasks vary widely, the objective remains the same: Perform a job with the highest possible quality in the shortest amount of time. 
For instance, to protect a car's underbody from corrosion, exposed welding seams are sealed by applying a polyvinyl chloride layer (PVC). The welding seams need to be traversed by a robot to apply the material.


\emph{Problem class and mathematical formulation:} The problem belongs to the domain of optimization problems, specifically to the class: traveling salesman problem, which is classified as NP-hard~\cite{TSP}. A typical mathematical formulation of the model is a weighted graph, which encodes the distance of all possible combinations of start- and endpoints. 
The goal is to find a combination of connected start- and endpoints that represent the shortest path that needs to be traversed, and therefore, the shortest time necessary. A mathematical formulation suitable for a quantum computer is Quadratic Unconstrained Binary Optimization (QUBO), which can be derived from the traveling salesman graph~\cite{Lucas_2014}. We assign binary variables to the decision of assigning a robot to a specified seam at a discretized time, for all seams and all times.
Hence, for a problem with $N$ seams, $2 \times N^2$ variables are needed to model the connection between two points on the graph \cite{9108020}. 

\emph{Algorithm:} The QUBO problem can then be solved using quantum annealing~\cite{DBLP:conf/kivs/MehtaMW17}. The most straightforward mapping requires $M$ qubits for $M$ binary variables. 
Another possible algorithm to solve this problem on a gate-based quantum computer is the Quantum Approximate Optimization Algorithm (QAOA) \cite{farhi2014quantum}.

\subsubsection{Vehicle Configuration Optimization}


\emph{Problem description:} Newly developed car components need to be tested before moving into series production. For this purpose, test vehicles are built. This application's objective is to ensure that every new component is at least built into a test vehicle once. Since many different components are developed each year, the efficient allocation of components to test vehicles is essential, ensuring that the number of built test vehicles is minimal. A powerful engine requires a corresponding transmission. 

Further, there might be dependencies to other vehicle options, e.\,g., an interior design options might depend on the chosen engine. This example shows that the complex dependencies between components can result in a complex, combinatorial problem. Thus, minimizing the number of test vehicles can become highly complex. Similar, configuration problems arise in other areas, e.\,g., vehicle sequencing, optimizing vehicle configurations for built-to-stock vehicles, and in other industries, e.\,g., the production of other consumer goods.

\emph{Problem class and mathematical formulation:}  This problem belongs to the class of boolean satisfiability problems (SAT). The different components are mapped to boolean variables, which are either $1$ (component is built in) or $0$ (component is not built in). Boolean clauses express the dependencies between the different components. The resulting mathematical formulation contains clauses with varying sizes, likely resulting in higher-order than quadratic terms (Polynomial Unconstrained Binary Optimization (PUBO)). 


\emph{Algorithm:} Even though it has been shown that SAT problems can also be mapped to QUBO problems, the reformulation of higher-order terms as quadratic functions leads to a high amount of ancillary variables. To avoid this increase of variables, the QAOA can, e.\,g., be used to solve the resulting PUBO problem \cite{Hadfield2018}.

\subsubsection{The relevance of industry benchmarks}

We showed how applications can be analyzed using a set of well-defined categories, in particular problem classes, mathematical formulation, and algorithms. Approximate solutions can be found using different quantum hardware systems. Though many different approaches are possible, the feasibility differs significantly depending on which formulation and hardware is applied. Progress in quantum hardware may not be directly translatable in progress in solving optimization problems. Thus, solving real-world reference problems provide a good proxy for evaluating progress.  
By deriving reference problems from different domains, the suitability of the underlying methods and the technology's maturity  can be evaluated. This approach allows the investigation of hardware characteristics for real-world industry applications. 
Robot path planning and vehicle configuration optimization are examples of two high-value use cases that can be generalized across many industrial applications beyond the automotive industry. For example, robots are increasingly used in all industry from consumer goods to airplanes.  Similarly, as products' customizability increases, configuration problems are emerging in several industries, e.g., the optimal configuration of IT systems. Therefore, industry application benchmarks can guide both academic and industry communities.
%


\section{Discussion}


We have identified 10+ application scenarios and 40+ use cases across the automotive value chain that are promising for mid- and long-term quantum-based methods, particularly in the domains of optimization, simulation, and machine learning. In these domains,  algorithms, such as quantum annealing, adiabatic optimization, hybrid algorithms for machine learning, and linear equations, that can evolve toward a quantum advantage on NISQ-era machines are emerging. Thus, it is instrumental to identify and pursue high-value problems that benefit from higher-quality solutions (e.\,g., solving more realistic, larger problem sizes) and higher performance. Solving these challenges can accelerate the development of quantum-enhanced solutions.

Industry applications are an essential factor in establishing ecosystems and accelerating the development of quantum computing. Currently, benchmark activities often focus on low-level aspects, such as gate fidelity. There is a lack of  application and end-to-end benchmarks that push development towards real-world problems. By describing these high-impact business problems, we can incentivize the exploration of novel solutions spaces, including hybrid and quantum-inspired solutions, and advance the commercial development of quantum computing, while also enhancing our understanding of the technology's potentials for our industry.

In the future, we will evolve the identified reference problems to benchmarks to create a benchmark suite for industry problems. For this purpose, we will develop reference implementations comprising configurable workloads and datasets with different complexity and defined metrics.  We postulate that comprehensive benchmark suites on every level are essential to guide progress toward high-value applications that provide a potential quantum advantage.  Further, standardized benchmarks developed based on these problems are essential to understand the current state of quantum computing (e.\,g., problem sizes, performance, scalability), illustrate best practices, and make predictions about the progress of quantum computing.

\end{document}